\documentclass[prl,twocolumn,superscriptaddress,showpacs,floatfix,longbibliography]{revtex4-1}
\usepackage{mathrsfs,braket}
\usepackage{amssymb, amsbsy, amsmath, latexsym, dsfont, array, layout,
graphicx,mathrsfs,color,bm}

\newcommand{\one}{\mathds{1}}

\begin{document}

\title{Observation of Non-Bloch Parity-Time Symmetry and Exceptional Points}

\author{Lei Xiao}
\thanks{These authors contributed equally to this work}
  \affiliation{Beijing Computational Science Research Center, Beijing 100084, China}

\author{Tianshu Deng}
\thanks{These authors contributed equally to this work}
  \affiliation{Institute for Advanced Study, Tsinghua University, Beijing, 100084, China}
\author{Kunkun Wang}
  \affiliation{Beijing Computational Science Research Center, Beijing 100084, China}

\author{Zhong Wang}\email{wangzhongemail@tsinghua.edu.cn}
  \affiliation{Institute for Advanced Study, Tsinghua University, Beijing, 100084, China}
\author{Wei Yi}\email{wyiz@ustc.edu.cn}
  \affiliation{CAS Key Laboratory of Quantum Information, University of Science and Technology of China, Hefei 230026, China}
  \affiliation{CAS Center For Excellence in Quantum Information and Quantum Physics, Hefei 230026, China}
\author{Peng Xue}\email{gnep.eux@gmail.com}
  \affiliation{Beijing Computational Science Research Center, Beijing 100084, China}

\begin{abstract}
Parity-time (\emph{PT})-symmetric Hamiltonians have widespread significance in non-Hermitian physics.
A \emph{PT}-symmetric Hamiltonian can exhibit distinct phases with either real or complex eigenspectrum, while the transition points in between, the so-called exceptional points, give rise to a host of critical behaviors that holds great promise for applications.
For spatially periodic non-Hermitian systems, \emph{PT} symmetries are commonly characterized and observed in line with the Bloch band theory, with exceptional points dwelling in the Brillouin zone. Here, in nonunitary quantum walks of single photons, we uncover a novel family of exceptional points beyond this common wisdom. These ``non-Bloch exceptional points'' originate from the accumulation of bulk eigenstates near boundaries, known as the non-Hermitian skin effect, and inhabit a generalized Brillouin zone. Our finding opens the avenue toward a generalized \emph{PT}-symmetry framework, and reveals the intriguing interplay between \emph{PT} symmetry and non-Hermitian skin effect.
\end{abstract}

\maketitle

While Hermiticity of Hamiltonians is a fundamental axiom in the standard quantum mechanics for closed systems, non-Hermitian Hamiltonians arise in open systems and possess unique features. Particularly, a wide range of non-Hermitian Hamiltonians, protected by the parity-time (\emph{PT}) symmetry, can have entirely real eigenvalues~\cite{Bender,El-Ganainy2018review,Ozdemir2019review,Miri2019review}.
A \emph{PT}-symmetric Hamiltonian generally has two phases, the exact \emph{PT} phase and the broken \emph{PT} one, with real and complex eigenenergies, respectively.
The transition points between these phases are called exceptional points, on which eigenstates and eigenenergies coalesce while the Hamiltonian becomes nondiagonalizable.
\emph{PT} symmetry and exceptional points are ubiquitous in non-Hermitian systems, and lead to dramatic consequences and promising applications such as unidirectional invisibility~\cite{Regensburger2012}, single-mode lasers~\cite{Feng2014,Hodaei2014}, enhanced sensing~\cite{Chen2017,Hodaei2017},
topological energy transfer~\cite{Xu2016}, and nonreciprocal wave propagation~\cite{Ruter,Peng2014}, to name just a few. In practice, physical systems with \emph{PT} symmetry are often based on spatially periodic structures (e.g., photonic lattices or microwave arrays)~\cite{El-Ganainy2018review,Makris}, where the notion of Bloch band greatly simplifies their characterization as each Bloch wave is treated independently.

Here we uncover a novel class of exceptional points beyond this Bloch-band picture in periodic systems. This work is partially stimulated by recent discoveries of non-Hermitian topological systems whose topological properties are highly sensitive to boundary conditions, in sharp contrast to their Hermitian counterparts. Specifically, for a generic family of non-Hermitian systems under the open-boundary condition (OBC), all eigenstates accumulate near the boundaries, whereas they always behave as extended Bloch waves under a periodic boundary condition (PBC).
This phenomenon, known as the non-Hermitian skin effect, invalidates the conventional bulk-boundary correspondence and necessitates a re-definition of topological invariants~\cite{Helbig,Xiao,Weidemann,Ghatak,Hofmann}. Whereas topological physics has been the focus in previous studies~\cite{wang2018a,wang2018b,Kunst,murakami,Lee2019,Kawabata2019,AVT18}, a fundamental question remains whether the non-Hermitian skin effect has significant consequences beyond topology, among which the interplay of non-Hermitian skin effect and \emph{PT} symmetry is arguably the most intriguing~\cite{Longhi1,Longhi2}. In this work, we experimentally observe
exceptional points generated by the non-Hermitian skin effect. Despite a translationally invariant bulk, the observed exceptional points exist in a ``generalized Brillouin zone'' (GBZ)~\cite{wang2018a,wang2018b,murakami,Longhi1,Longhi2} (rather than the standard Brillioun zone), thus representing an unexplored
class of exceptional points beyond the conventional Bloch-band framework. Just as the framework with conventional Bloch bands has been commonly adopted to describe periodic lattices in physical systems ranging from condensed matter to photonics, the generalized mechanism of \emph{PT} symmetry, confirmed by our observation of ``non-Bloch exceptional points'', is relevant to a broad class of non-Hermitian platforms with periodic structures.

\begin{figure*}
\centering
\includegraphics[width=0.8\textwidth]{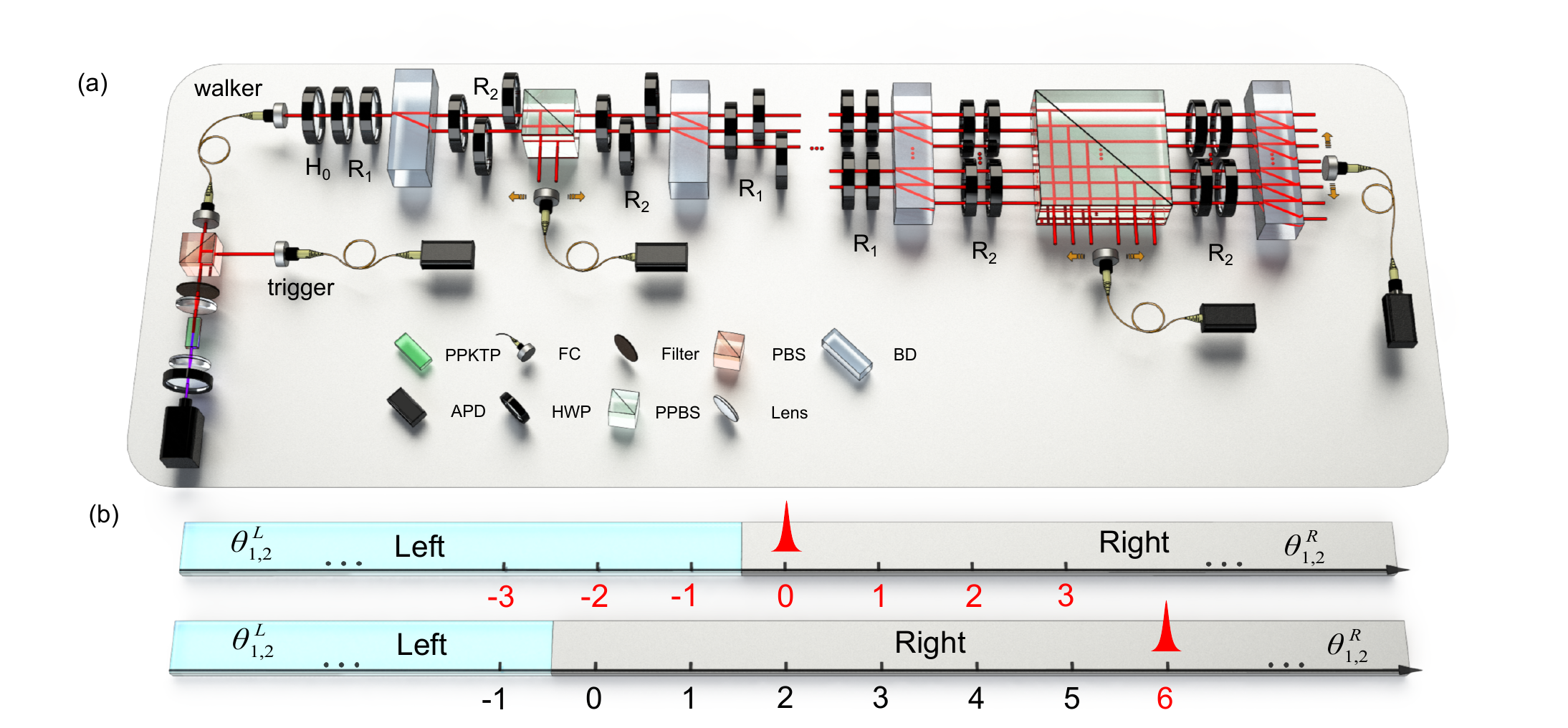}
\caption{Experimental implementation.
(a) A pair of photons is generated via the spontaneous parametric down conversion in the periodically poled potassium titanyl phosphate crystal (PPKTP), with one serving as a trigger and the other (walker) projected into the quantum-walk network as the walker photon. After passing through a polarizing beam splitter (PBS) and a half-wave plate (HWP), the polarization of the walker photon is initialized as $|0\rangle$. It then undergoes a quantum walk through an interferometric network, composed of HWPs, beam displacers (BDs), and partially polarizing beam splitter (PPBS), and is finally detected by avalanche photodiodes (APDs), in coincidence with the trigger photon.
(b) The domain-wall geometry of non-Hermitian quantum walks. Upper panel (scheme I): the walker starts near the domain wall at $x=0$. Lower panel (scheme II): the walker starts from the bulk (i.e., far away from the domain wall) position $x=6$.}
\label{fig:fig1}
\end{figure*}

\begin{figure*}
\includegraphics[width=\textwidth]{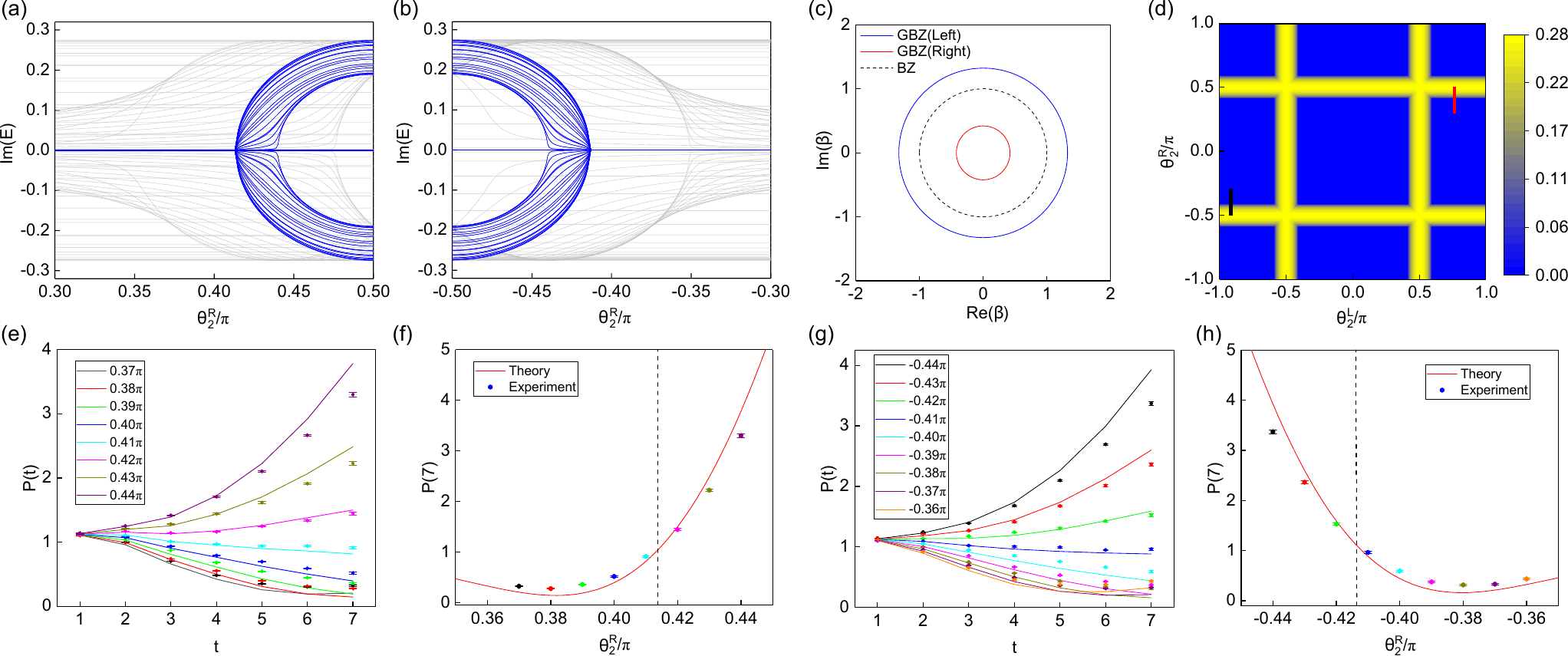}
\caption{Non-Bloch exceptional points from domain-wall measurements.
(a, b) Imaginary part of quasienergies, Im$(E)$ versus $\theta_2^R$. Other coin parameters are fixed at $\theta_1^R=0.5625\pi$ and $\theta_1^L=-0.0625\pi$. We take $\theta_2^L=0.75\pi$ and $-0.9735\pi$ for (a) and (b), respectively. Blue and gray lines represent the OBC (non-Bloch) and PBC (Bloch) spectra, respectively. The blue non-Bloch spectra feature an exceptional point at $|\theta_2^R|=0.413\pi$, while the gray Bloch spectra remain complex-valued throughout.
(c) Brillouin zone and GBZ for $\theta^R_1=0.5625\pi$, $\theta^L_1=-0.0625\pi$, $\theta_2^R=-0.44\pi$, and $\theta_2^L=-0.9375\pi$.
(d) Numerically calculated $\text{max}[\text{Im}(E)]$ for $\theta^R_1=0.5625\pi$ and $\theta^L_1=-0.0625\pi$. The yellow (blue) region is the broken (exact) \emph{PT} phase. The red and black cuts correspond to (a, e, f) and (b, g, h), respectively.
(e) Experimentally measured $P(t)$ (symbols) with an initial state $\ket{\psi(t=0)}=\ket{0}_x\otimes\ket{0}_\text{coin}$ up to seven steps for eight different values of $\theta^R_2$, together with the theoretical predictions (curves). Other coin parameters are the same as in (a), e.g., $\theta_2^L=0.75\pi$.
(f) $P(t=7)$ under the same parameters as those in (c). Error bars indicate the statistical uncertainty obtained by assuming Poissonian statistics. The red line is plotted from numerical simulations of seven-step evolutions, from which the exceptional point is predicted to be $\theta_2^R=0.413\pi$ [by requiring $P(7)=1$], consistent with the theoretical prediction from Eq.~(\ref{EP}).
(g, h) The same as (e, f) except that $\theta_2^L=-0.9735\pi$. From the numerical simulations (red line), the exceptional point in (h) is at $\theta_2^R=-0.412\pi$.
The loss parameter is fixed at $\gamma=0.2746$ throughout our experiment.}
\label{fig:fig2}
\end{figure*}

In general, a discrete-time, nonunitary quantum walk can be characterized by $|\psi(t)\rangle=U^t|\psi(0)\rangle$ ($t=0,1,2,\cdots$), which amounts to a stroboscopic simulation of the time evolution with initial state $|\psi(0)\rangle$, and generated by the non-Hermitian effective Hamiltonian
$H_{\rm eff}$ with $U:=e^{-iH_{\rm eff}}$. To be concrete, we take the following one-dimensional Floquet operator
\begin{equation}
U=R\left(\frac{\theta_1}{2}\right)S_2R\left(\frac{\theta_2}{2}\right)M R\left(\frac{\theta_2}{2}\right)S_1R\left(\frac{\theta_1}{2}\right),
\label{U}
\end{equation}
where the shift operators, $S_{1} =\sum_{x}|x\rangle\langle x|\otimes| 0\rangle\langle 0|+| x+1\rangle\langle x|\otimes| 1\rangle\langle 1|$ and $
S_{2}  =\sum_{x}|x-1\rangle\langle x|\otimes| 0\rangle\langle 0|+| x\rangle\langle x|\otimes| 1\rangle\langle 1|$, selectively shift the walker along a one-dimensional lattice (with lattice sites labeled by $x$) in a direction that depends on its internal coin state $|0\rangle$($|1\rangle$), the $+1$ ($-1$) eigenstate of the Pauli matrix $\sigma_z$. The coin operator $R(\theta) =\one_w \otimes e^{-i \theta \sigma_{y}}$, with $\one_w=\sum_x|x\rangle\langle x|$, rotates coin states without shifting the walker position. The gain and loss is implemented by $M=\one_\text{w} \otimes e^{\gamma\sigma_z}$.

We implement the quantum walk using a single-photon interferometric network [Fig.~\ref{fig:fig1}(a)], where
coin states $|0\rangle$ and $|1\rangle$ are encoded in the horizontal and vertical photon polarizations, respectively. Rotations of the coin states ($R$) are implemented by HWPs. Shift operators $S_{1,2}$ are realized by beam displacers that allow the transmission of vertically polarized photons while displacing horizontally polarized ones into neighboring positions.
Finally, the gain/loss is implemented by a PPBS, which reflects state $\ket{1}$ with a probability $p$, and directly transmits state $\ket{0}$.
Thus, the PPBS realizes $M_E=\one_\text{w}\otimes \left(|0\rangle\langle 0|+\sqrt{1-p}|1\rangle\langle 1|\right)$, which is related to $M$ as $M=e^\gamma M_E$, with $\gamma=-\frac{1}{4}\ln (1-p)$. We therefore readout $|\psi(t)\rangle$ from our experiment with $M_E$ by adding a factor $e^{\gamma t}$. More details of our experimental setup can be found in ~\cite{supp}.

Under the Floquet operator $U$, the directional hopping in $S_{1,2}$ and the gain/loss in $M$ conspire to generate non-Hermitian skin effect~\cite{Xiao}. When a domain wall is created between two regions with different parameters, e.g., $\theta^L_{1,2}$ and $\theta^R_{1,2}$ for the left and right regions in Fig.~\ref{fig:fig1}(b), all the eigenstates of $U$ are localized at the domain wall~\cite{Xiao}. While the non-Hermitian skin effect dramatically affects topological properties, here we focus on the impact of non-Hermitian skin effect on the emergence of \emph{PT} symmetry and exceptional points.

\begin{figure*}
\includegraphics[width=\textwidth]{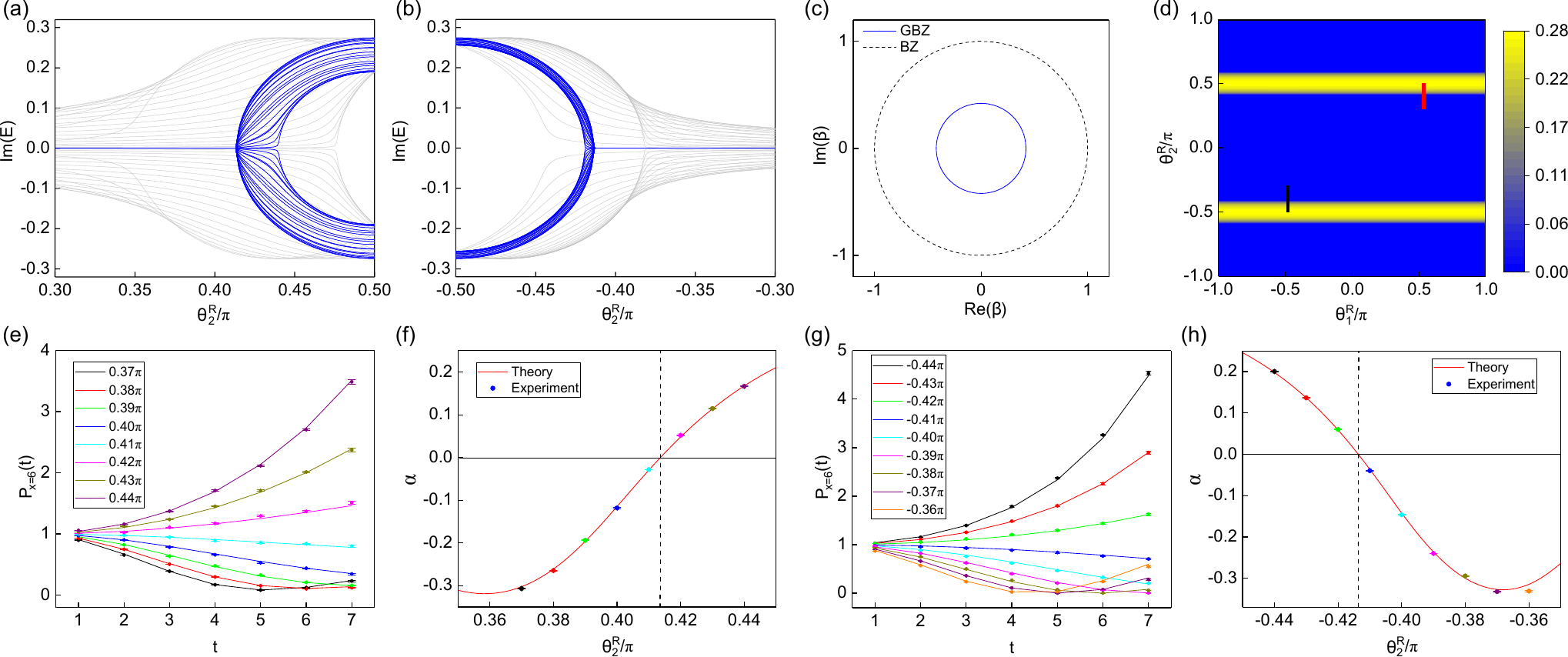}
\caption{Non-Bloch exceptional points from bulk measurements.
(a) Im$(E)$ versus $\theta_2^R$, with $\theta_1^R=0.5625\pi$ fixed.
(b) Im$(E)$ versus $\theta_2^R$, with $\theta_1^R=-0.4688\pi$ fixed.
(c) GBZ of the right region, for $\theta^R_1=-0.4688\pi$ and $\theta_2^R=-0.44\pi$.
(d) Numerically calculated max[Im($E$)]. The blue and yellow regions correspond to the exact and broken \emph{PT} phase, respectively.
(e) Experimentally measured $P_{x=6}(t)$ (symbols) with an initial state $\ket{\psi(0)}=\ket{6}_x\otimes\ket{0}_\text{coin}$ for eight values of $\theta^R_2$, compared to the theoretical predictions (curves). Here we fix $\theta^R_1=0.5625\pi$.
(f) Exponent $\alpha$ versus $\theta_2^R$ extracted from the data in (e). The red line is plotted from numerical simulations of seven-step evolutions, from which the exceptional point is predicted to be $\theta_2^R=0.414\pi$ (by requiring $\alpha=0$), consistent with the theoretical prediction from Eq.~(\ref{EP}).
(g, h) The same as (e, f), except that $\theta^R_1=-0.5\pi$. (e, f) and (g, h) correspond to the red and black cuts in the phase diagram (d), respectively. From the numerical simulation (red line), the exceptional point in (h) is at $\theta_2^R=-0.414\pi$.
The left region is idle throughout measurements, which are performed in the right region only. Without loss of generality, we take $\theta^L_1=-0.0625\pi$ and $\theta_2^L=0.75\pi$ for (e, f), and $\theta^L_1=-0.0625\pi$ and $\theta_2^L=-0.9375\pi$ for (g, h).
}
\label{fig:fig3}
\end{figure*}

The exact (broken) \emph{PT} phase corresponds to the absence (presence) of nonzero imaginary parts in the eigen spectrum (quasienergies) of $H_\text{eff}$. Note that we use these terms in a general sense, including pseudo-Hermiticity whose role is, regarding spectral reality, similar to the original \emph{PT} symmetry~\cite{Mostafazadeh,Mostafazadeh2}. In Figs.~\ref{fig:fig2}(a, b), we show in blue the calculated imaginary parts of quasienergies, Im($E$), for the domain-wall geometry with OBC at the two ends [see Fig.~\ref{fig:fig1}(b)]. For both Figs.~\ref{fig:fig2}(a, b), an exceptional point is found as $\theta_2^R$ is varied while fixing other parameters. Remarkably, the exceptional point cannot be deduced from the Bloch band theory. The Bloch theory suggests that the continuous bulk spectra of $U$ under the domain-wall geometry are the union of the spectra corresponding to the left and right bulks, which are respectively obtained from the Bloch Floquet operator $U(k)$ ($k\in[0,2\pi]$, i.e., within the standard Brillioun zone) associated with the left (with parameters $\theta_{1,2}^L$) and right (with $\theta_{1,2}^R$) bulk.
These spectra are shown in gray in Figs.~\ref{fig:fig2}(a, b), which dramatically differ from the actual (non-Bloch) spectra under the domain-wall geometry.

This discrepancy is due to the aforementioned non-Hermitian skin effect. The exponential decay of eigenstates in the real space means that the Bloch phase factor $e^{ik}$, which corresponds to extended plane waves, should be replaced by a factor $\beta$ ($|\beta|\neq 1$ in general) in order to generate the eigenspectra under the OBC. Furthermore, $\beta$ must belong to a closed loop in the complex plane, dubbed the GBZ~\cite{wang2018a,murakami}, which typically deviates from the unit circle (Fig.~\ref{fig:fig2}(c)). For $\beta\in$GBZ, eigenenergies under the OBC are recovered by performing the analytic continuation $U(k)|_{  e^{ik}\rightarrow\beta}$, and taking the logarithm of eigenvalues of $U(\beta)$.
Crucially, we find that $U(\beta)$ satisfies the $\eta$-pseudo-unitarity
\begin{align}
\eta U^{-1}(\beta)\eta^{-1}=U^{\dagger}(\beta)\big|_{\beta\in\text{GBZ}}, \label{eq:Usymm}
\end{align}
when $|\cos\theta_{2}^{L(R)}|>|\tanh\gamma|$~\cite{supp}. Here $\eta=\sum_n|\chi_n\rangle\langle\chi_n|$, where $\{|\chi_n\rangle\}$ is the collection of left eigenstates of $U(\beta)$. Equation (\ref{eq:Usymm}) corresponds to the $\eta$-pseudo-Hermiticity of the non-Hermitian effective Hamiltonian: $\eta H_\text{eff}(\beta)\eta^{-1} = H^\dag_\text{eff}(\beta)$~\cite{Mostafazadeh,Mostafazadeh2}, which is a generalization of the \emph{PT} symmetry, and guarantees the reality of quasienergies as long as the relation holds. As such, the GBZ theory predicts non-Bloch exceptional points at
\begin{align}
|\cos\theta_2^{L(R)}|=|\tanh\gamma|.
\label{EP}
\end{align}

We observe exceptional points by probing probabilities of the photon surviving in the quantum walk after each time step $t$, which are constructed from photon-number measurements up to $t$~\cite{supp}. They are then multiplied by a factor $e^{2\gamma t}$ (due to the aforementioned difference between $M_E$ and $M$) to yield the corrected probability $P(t)$ that corresponds to the wave function norm, whose long-time behavior is $P(t)=\bra{\psi(t)}\psi(t)\rangle\sim e^{2\text{max}[\text{Im}(E)]t}$.
Therefore, an exponential growth of $P(t)$ indicates the broken \emph{PT} phase. By contrast, $P(t)$ in the exact \emph{PT} phase typically approaches a steady-state value of order of unity. Such a feature enables us to extract the location of exceptional points by tracking the time evolution of the corrected probability. Experimentally, this is achieved through two schemes: (I) the domain wall scheme and (II) the bulk scheme.

In the first scheme, we initiate the photon walker near the domain wall, as illustrated in the upper panel of Fig.~\ref{fig:fig1}(b), with the initial state $\ket{\psi(t=0)}=\ket{0}_x\otimes\ket{0}_\text{coin}$. We then measure the corrected probability along the red and black cuts in the numerically simulated phase diagram [Fig.~\ref{fig:fig2}(d)], where the blue and yellow regions correspond to the exact and broken \emph{PT} phase, respectively.
In Fig.~\ref{fig:fig2}(e) (red cut), $P(t)$ grows with $t$ for $\theta_2^R\geq 0.42\pi$ and decreases for $\theta_2^R\leq 0.41\pi$.
Therefore, we infer the presence of an exceptional point between $\theta_2^R=0.41\pi$ and $0.42\pi$. This is consistent with Eq.~(\ref{EP}), which predicts an exceptional point at $\theta_2^R=\pm 0.413\pi$. We arrive at the same conclusion by measuring corrected probabilities at the time step $t=7$ (Fig.~\ref{fig:fig2}(f)), which become prominently larger than $1$ in the broken \emph{PT} phase. Similarly, Figs.~\ref{fig:fig2}(g, h) (blue cut) indicate an exceptional point in the region $\theta^R_2\in\left[-0.42\pi,-0.41\pi\right]$, again consistent with Eq.~(\ref{EP}).

The second scheme is based on local measurements in the bulk. The walker starts from a position $x=x_0$ far from the domain wall [Fig.~\ref{fig:fig1}(b), lower panel], and the subsequent corrected probability at $x=x_0$, i.e.,  $P_{x_0}(t)= |\bra{0}\otimes\bra{x_0}\psi(t)\rangle |^2+ |\bra{1}\otimes\bra{x_0}\psi(t)\rangle|^2$, is measured. In the broken \emph{PT} phase,
the corrected probability grows as $P_{x_0}(t)\propto e^{\alpha t}$, where $\alpha>0$ is given by the imaginary part of quasienergies at certain special points of the GBZ~\cite{Longhi1,Longhi2}. In the exact \emph{PT} phase, by contrast, $P_{x_0}(t)$ features an oscillatory behavior at short times, enveloped by an overall decay characterized by $\alpha<0$ (which approaches zero as the evolution time increases)~\cite{Longhi2,supp}.
In our experiment, we fix $x_0=6$ in the right region, leaving the left region idle. The imaginary parts of quasienergy spectra under OBC are plotted in Figs.~\ref{fig:fig3}(a, b), along the red and black cuts in Fig.~\ref{fig:fig3}(d), respectively. The spectra are calculated by diagonalizing $U(\beta)|_{\beta\in\text{GBZ}}$ for the right region, with GBZ shown in Fig.~\ref{fig:fig3}(c). Along the red cut ($\theta^R_1=0.5625\pi$), the measured $P_{x=6}(t)$ exhibits growth for $\theta_2^R\geq 0.42\pi$, and decreases for $\theta_2^R\leq 0.41\pi$ as illustrated in Fig.~\ref{fig:fig3}(e), indicating an exceptional point within $\left[0.41\pi,0.42\pi\right]$. This is consistent with Fig.~\ref{fig:fig3}(a). Moreover, we fit $P_{x=6}(t)$ exponentially in Fig.~\ref{fig:fig3}(f). While the accuracy in $\alpha$ is limited by the small number of experimentally feasible steps~\cite{supp}, the fitted $\alpha$ does yield qualitatively consistent results: the sign of $\alpha$ is positive (negative) in the broken (exact) \emph{PT} phase. A similar exceptional point is observed along the black cut ($\theta^R_1=-0.5\pi$) in Figs.~\ref{fig:fig3}(g, h).

Notably, under the bulk scheme, we are able to establish non-Bloch \emph{PT} symmetry and detect non-Bloch exceptional points from dynamics purely in the bulk, i.e., essentially under PBC. This highlights the observed non-Bloch exceptional points as intrinsic properties of our system, rather than mere finite-size effects. While we have revealed the non-Bloch \emph{PT} transition using a seven-step quantum walk, alternative designs (such as the time-multiplexed framework~\cite{CDQ+18}) with the potential of achieving a longer evolution time would enable a more precise probe of the transition, including accurate determination of the Lyapunov exponent~\cite{Longhi1,Longhi2}.

The significance of the observed non-Bloch \emph{PT} symmetry and exceptional points is further enhanced by the following generic finding: in the presence of non-Hermitian skin effect, the Bloch energy spectra (calculated from the Brillouin zone) can never have \emph{PT} symmetry. In fact, recent theoretical works prove that, if a system features non-Hermitian skin effect under the OBC, the associated Bloch spectra must form loops in the complex plane~\cite{Zhangkai,Okuma,LLTG20}. However, looplike spectra cannot lie in the real axis, thus forbidding entirely real spectrum.
In sharp contrast, the non-Bloch spectra calculated from the GBZ, which correctly reflect eigenenergies under the experimentally relevant OBC, form arcs or lines enclosing no area. Real spectra and \emph{PT} symmetry are henceforth enabled. Therefore, non-Bloch \emph{PT} symmetry is the only general mechanism for achieving \emph{PT} symmetry in the presence of non-Hermitian skin effect.

The observed interplay between non-Hermitian skin effect and \emph{PT} symmetry underlines a fundamentally new mechanism for \emph{PT} symmetry and exceptional points in periodic systems, and demonstrates the power of non-Bloch band theory beyond topology. Since both the non-Hermitian skin effect and \emph{PT} symmetry are generic features of a large class of non-Hermitian systems, the mechanism is general and applies to a variety of non-Hermitian platforms ranging from photonic lattices to cold atoms.
In view of the potential utilities of exceptional points, the non-Bloch exceptional points observed here would inspire novel designs and applications such as enhanced sensing with interface-sensitive, ultrahigh spatial resolutions~\cite{Weidemann,Hodaei2017,Chen2017}, or
robust wireless power transfer that are tunable by the interface geometry~\cite{AYF17}.

\begin{acknowledgments}
This work has been supported by the National Natural Science Foundation of China (Grant No. 12025401, No. 11674189, No. U1930402 and No. 11974331) and a start-up fund from the Beijing Computational Science Research Center. W.Y. acknowledges support from the National Key Research and Development Program of China (Grants No. 2016YFA0301700 and No. 2017YFA0304100).  K. W. and L. X.  acknowledge support from the Project Funded by China Postdoctoral Science Foundation (Grants No. 2019M660016 and No. 2020M680006).
\end{acknowledgments}

%L. X. and T. D. contributed equally to this work.

\bibliography{reference}

\newpage
\begin{widetext}
\appendix

\renewcommand{\thesection}{\Alph{section}}
\renewcommand{\thefigure}{S\arabic{figure}}
\renewcommand{\thetable}{S\Roman{table}}
\setcounter{figure}{0}
\renewcommand{\theequation}{S\arabic{equation}}
\setcounter{equation}{0}

\section{Supplemental Material for ``Observation of non-Bloch parity-time symmetry and exceptional points''}

\subsection{Experimental Methods}
As illustrated in Fig. 1(a) and discussed in the main text, coin states are encoded in the photon polarizations, with $|0\rangle$ and $|1\rangle$ corresponding to the horizontal and vertical polarizations, respectively. The walker states are represented by the spatial modes of photons, with the lattice sites labelled by $x$. The walker photon is initialized at either the domain wall ($x=0$) or at a site far from the domain wall ($x=x_0$), and is projected onto one of the polarization states $\ket{0}$ by a polarizing beam splitter (PBS) and a half-wave plate (HWP). While the coin operator $R$ is implemented by HWPs, the shift operator $S_1$ ($S_2$) is realized by a beam displacer (BD), which allows the direct transmission of vertically polarized photons and displaces horizontally polarized photons laterally to a neighboring spatial mode.

Non-unitarity is introduced by photon loss, which is realized by a mode-selective loss operator
\begin{equation}
M_E=\one_\text{w}\otimes \left(|0\rangle\langle 0|+\sqrt{1-p}|1\rangle\langle 1|\right)
\end{equation}
with $\one_\text{w}=\sum_x |x\rangle\langle x|$, realizing a partial measurement at every time step. The loss operator $M_E$ is implemented by a partially polarizing beam splitter (PPBS). In our experiment, $M_E$, rather than $M=e^\gamma M_E$ [$\gamma=-\frac{1}{4}\ln (1-p)$], is implemented. Nevertheless, the experimentally implemented $t$-step quantum-walk dynamics can be mapped to that under $U$ by multiplying a time-dependent factor $e^{\gamma t}$. Under the Floquet operator $U$, the time-evolved state is given by $|\psi(t)\rangle=U^t|\psi(0)\rangle$ ($t=0,1,2,\cdots$). Therefore, a quantum walk stroboscopically simulates the nonunitary time evolution generated by the non-Hermitian effective Hamiltonian $H_{\rm eff}$, where $U:=e^{-iH_{\rm eff}}$. Typical eigen spectra of $H_{\rm eff}$ are shown in Figs.~2(b) and 3(b).

In the first scheme of detecting non-Bloch exceptional points, the walker starts from $x=0$ shown in Fig.~1(b). We measure the corrected probability after $t$-step quantum-walk dynamics $P(t)=\bra{\psi(t)}\psi(t)\rangle$ with initial state $\ket{\psi(0)}=\ket{0}\otimes\ket{0}$ through photon-number measurements, which are registered by the coincidences between one of the avalanche photodiodes (APDs) in the detection stage and that for the trigger photon. The corrected probability $P(t)$ can be calculated from the photon-number measurements multiplied (i.e., corrected) by a time-dependent factor $e^{2\gamma t}$
\begin{equation}
P(t)=e^{2\gamma t}\sum_x \frac{N(t,x)}{\sum_{x}\left[N(t,x)+\sum_{t'=1}^{t} N_L(t',x)\right]},
\end{equation}
where $N(t,x)$ is the number of the photons detected at $x$ after a $t$-step evolution and $N_L(t,x)$ is the photon loss caused by the partial measurement $M_E$ at the time step $t$.

For the second scheme in which the walker starts from $x=x_0$ far from the domain wall, we measure the corrected probability at $x=x_0$ after a $t$-step quantum walk $P_{x_0}(t)=\Big|\bra{0}\otimes\bra{x_0}\psi(t)\rangle\Big|^2+\Big|\bra{1}\otimes\bra{x_0}\psi(t)\rangle\Big|^2$, with initial state $\ket{\psi(0)}=\ket{x_0}\otimes\ket{0}$. The expression for $P_{x_0}(t)$ is therefore
\begin{equation}
P_{x_0}(t)=e^{2\gamma t}\frac{N(t,x_0)}{\sum_x\left[N(t,x)+\sum_{t'=1}^t N_L(t',x)\right]}.
\end{equation}

If all the eigenenergies of the system are real, i.e., the system is in the exact parity-time (\emph{PT}) phase, $P(t)$ approaches a steady-state value at long times, while $P_{x_0}(t)$ decays towards $0$ by a power law~\cite{sLonghi1}. In contrast, after the system crosses the exceptional point and becomes \emph{PT} broken, quasienergies are typically complex. $P_{x_0}(t)$ and $P(t)$ would then increase exponentially with time.

To demonstrate the exponential dependence of $P_{x_0}(t)$ in time (for the \emph{PT} broken phase), we numerically fit $P_{x_0}(t)$ for quantum-walk dynamics up to $t=150$ steps, with either exponential or power-law dependence in time. We show the accumulated variance $\sum_t \left[P_{x=6}(t)-f(t)\right]^2/f^2(t)$ in Extended Data Fig.~\ref{fig:figS1} across the exceptional points, where $f(t)$ is the fitting function. Apparently, the errors with the exponential fit is always smaller in the \emph{PT} broken phase, and larger in the exact \emph{PT} phase. This justifies the exponential fit in Fig.~3 of the main text.

\subsection{Quasienergy spectrum under different boundary conditions}

The Floquet operator $U$ defined in the main text features the non-Hermitian skin effect, which can be revealed by plotting its quasienergy spectra on the complex plane, under both the periodic and open boundary conditions. As illustrated in Fig.~\ref{fig:figS2}, the quasienergy spectrum forms a closed loop on the complex plane under the periodic boundary condition (i.e. it has a nontrivial point-gap topology), but becomes lines or arcs under the open boundary condition. As proved in~\cite{sZhangkai,sOkuma}, eigenspectrum forming loops under the periodic boundary condition is closely related to the non-Hermitian skin effect. More important for our study, real spectra and \emph{PT} symmetry only become possible under the open boundary condition, which makes the non-Bloch \emph{PT} symmetry the only general mechanism for achieving \emph{PT} symmetry in the presence of non-Hermitian skin effect.

\begin{figure*}
\centering
\includegraphics[width=0.65\textwidth]{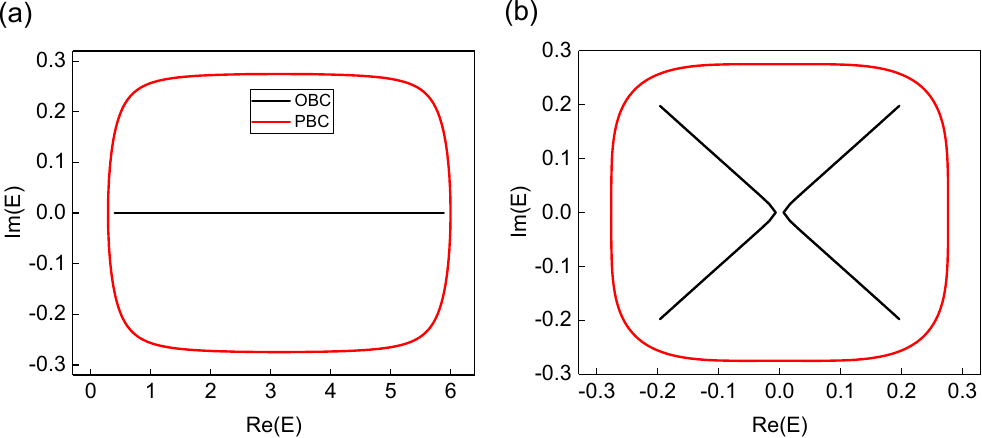}
\caption{Quasienergy spectra on the complex plane, under the open (black) and periodic (red) boundary conditions. We choose the same parameters as those in Fig.~2(c) of the main text, with $\theta_1^R=0.5625\pi$, $\theta_2^R=-0.44\pi$, $\theta_1^L=-0.0625\pi$, $\theta_2^L=-0.9375\pi$, and $\gamma = 0.2746$. Spectra of the left and right bulks are shown in (a) and (b), respectively.
}
\label{fig:figS2}
\end{figure*}

\subsection{Generalized Brillouin zone, non-Bloch \emph{PT} symmetry, and non-Bloch exceptional points}

In this section, we first derive the generalized Brillouin zone (GBZ). The analytic continuation of the Bloch Hamiltonian (or Floquet operator) to the GBZ yields the actual quasienergy spectra for the experimentally relevant open-boundary systems. We then show the non-Bloch \emph{PT} symmetry, i.e., \emph{PT} symmetry and exceptional points that exist in the GBZ rather than the conventional Brillouin zone (BZ). We will derive the analytic formula for the non-Bloch exceptional points given in the main text: $|\cos\theta_{2}|=|\tanh\gamma|$.

\begin{figure*}
\centering
\includegraphics[width=0.65\textwidth]{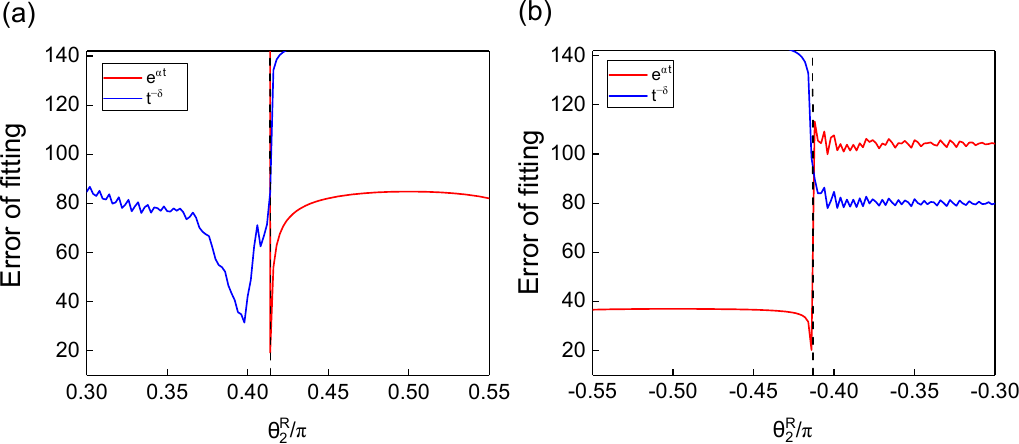}
\caption{Errors for fitting the corrected probability.
The accumulated variance $\sum_t \left[P_{x=6}(t)-f(t)\right]^2/f^2(t)$ for an exponential fit (red) with $f(t)=e^{\alpha t}$; and a power-law fit (blue) with $f(t)=t^{-\delta}$. Here $\alpha$ and $\delta$ are fitting parameters. Parameters in (a) and (b) are the same as those in Figs.~3(f) and (h), respectively.
}
\label{fig:figS1}
\end{figure*}

\subsubsection{Generalized Brillouin zone}
We start from the Floquet operator $U$, which can be decomposed as $U=FMG$, where
\begin{align}
&F=R\left (\frac{\theta_{1}}{2}\right ) S_{2} R\left(\frac{\theta_{2}}{2}\right ), \\
&G=R\left(\frac{\theta_{2}}{2}\right ) S_{1} R\left(\frac{\theta_{1}}{2}\right ),\\
&M=\one_\text{w} \otimes e^{\gamma\sigma_z}.
\end{align}
Here, the coin operator $R$ and the shift operator $S_{1,2}$ read
\begin{align}
&R(\theta) =\one_\text{w} \otimes R_c(\theta), \\
&S_{1} =\sum_{x}|x\rangle\langle x|\otimes| 0\rangle\langle 0|+| x+1\rangle\langle x|\otimes| 1\rangle\langle 1|, \\
&S_{2} =\sum_{x}|x-1\rangle\langle x|\otimes| 0\rangle\langle 0|+| x\rangle\langle x|\otimes| 1\rangle\langle 1|,
\end{align}
with $\one_\text{w}=\sum_x|x\rangle\langle x|$ and $R_c(\theta)=e^{-i \theta \sigma_{y}}$. Since $\theta_1(x)$ and $\theta_2(x)$ are both constants in the bulk (i.e., away from the domain wall), we rewrite the Floquet operator as
\begin{align}
U=\sum_{x}|x\rangle\left\langle x+1\left|\otimes A_{m}+\right| x\right\rangle\left\langle x-1\left|\otimes A_{p}+\right| x\right\rangle\langle x| \otimes A_{s},
\end{align}
where
\begin{align}
&A_{m}=R_c\left(\frac{\theta_{1}}{2}\right)P_{0}R_c\left(\frac{\theta_{2}}{2}\right)M_c R_c\left(\frac{\theta_{2}}{2}\right)P_{0}R_c\left(\frac{\theta_{1}}{2}\right),\\
&A_{p}=R_c\left(\frac{\theta_{1}}{2}\right)P_{1}R_c\left(\frac{\theta_{2}}{2}\right)M_c R_c\left(\frac{\theta_{2}}{2}\right)P_{1}R_c\left(\frac{\theta_{1}}{2}\right),\\
&A_{s}=R_c\left(\frac{\theta_{1}}{2}\right)P_{1}R_c\left(\frac{\theta_{2}}{2}\right)M_c R_c\left(\frac{\theta_{2}}{2}\right)P_{0}R_c\left(\frac{\theta_{1}}{2}\right) +R_c\left(\frac{\theta_{1}}{2}\right)P_{0}R_c\left(\frac{\theta_{2}}{2}\right)M_c R_c\left(\frac{\theta_{2}}{2}\right)P_{1}R_c\left(\frac{\theta_{1}}{2}\right)
\end{align}
with $M_c=e^{\gamma\sigma_z}$, $P_0=|0\rangle\langle0|$ and $P_1=|1\rangle\langle1|$.
Following the standard approach of calculating the GBZ~\cite{swang2018a,smurakami}, we write down the general eigenstate of $U$ as
\begin{align}
\left|\psi\right\rangle=\sum_{x,j} \beta_{j}^{x}|x\rangle \otimes\left|\phi_{j}\right\rangle_{c},
\end{align}
where $|\phi_{j}\rangle_c$ is the coin state and $\beta_j$ is the spatial-mode function. From the eigen equation $U|\psi\rangle=\lambda|\psi\rangle$, we obtain the bulk eigen equation
\begin{align}
\label{Ambeta}
\left(A_{m} \beta+\frac{A_{p}}{\beta}+A_{s}-\lambda\right)\left|\phi\right\rangle_{c}=0.
\end{align}
Eq.~(\ref{Ambeta}) supports nontrivial solutions only when
\begin{align}
\label{detAmAp}
{\rm{det}} \left[A_{m} \beta+A_{p}\frac{1}{\beta}+A_{s}-\lambda\right]=0,
\end{align}
%Eq.~(\ref{detAmAp}) appears to be a quartic equation of $\beta$, however, it is in fact quadratic because, as a consequence of $\det{A_m}=\det{A_p}=\det{A_s}=1$ the coefficients of $\beta^2$ and $\beta^{-2}$ vanish.
which is a quadratic equation of $\beta$ because $\det A_{m}=\det A_p=0$. Explicitly, it reads

\begin{align}
\label{betaqua}
&-\beta^{2}\left(\cosh\gamma\cos\theta_{1}\cos\theta_{2}+\sinh\gamma\cos\theta_{1}\right)
+\left(\lambda+\frac{1}{\lambda}+2\cosh\gamma\sin\theta_{1}\sin\theta_{2}\right)\beta +\sinh\gamma\cos\theta_{1}-\cosh\gamma\cos\theta_{1}\cos\theta_{2}\nonumber\\&=0.
\end{align}
Equation (\ref{betaqua}) has two solutions denoted as $\beta_1(\lambda)$ and $\beta_2(\lambda)$. In the thermodynamic limit, the open-boundary condition (OBC) requires that~\cite{swang2018a,smurakami} \begin{equation} |\beta_1(\lambda)|=|\beta_2(\lambda)|, \end{equation}
which is the GBZ equation. Combining it with quadratic Eq.~(\ref{betaqua}), we have
\begin{align}
\label{beta12}
\left|\beta_{1}\right|=\left|\beta_{2}\right|=\sqrt{\left|\frac{\cosh \gamma \cos \theta_{2}-\sinh \gamma}{\cosh \gamma \cos \theta_{2}+\sinh \gamma}\right|}.
\end{align}
Therefore, the GBZ is a circle with radius
\begin{align}
\label{radius}
\left|\beta\right|=\sqrt{\left|\frac{\cosh \gamma \cos \theta_{2}-\sinh \gamma}{\cosh \gamma \cos \theta_{2}+\sinh \gamma}\right|}.
\end{align}
This GBZ can be parameterized as $\beta=|\beta|e^{ip}$, with $p\in\left[0,2\pi\right]$.  For the left and right bulk (see the main text), $\theta_2=\theta_2^L$ and $\theta_2^R$, respectively. Note that $\theta_1$ does not affect the GBZ.

\subsubsection{$\eta$-pseudo-unitarity and non-Bloch exceptional points}.
In the exact \emph{PT} phase, the Floquet operator must be pseudo-unitary, i.e., $\eta U^{-1}\eta^{-1}=U^\dag$, with $\eta$ being a Hermitian, invertible, linear operator~\cite{sMostafazadeh,sMostafazadeh2}. A key finding of our work is that, for non-Hermitian systems with non-Hermitian skin effect, the pseudo-unitarity of $U$ is not in the BZ, but in the GBZ. In other words, we cannot find an $\eta$ such that $\eta U^{-1}(k)\eta^{-1}\neq U^{\dagger}(k)$ with $k\in\left[0,2\pi\right]$; however, such a symmetry can exist in the GBZ: \begin{equation} \eta U(\beta)^{-1}\eta^{-1}=U^{\dagger}(\beta)\big|_{\beta\in\text{GBZ}}. \label{pseudo-Supp} \end{equation}
Here, $U(\beta)$ is defined as the analytic continuation of the Bloch Floquet operator $U(k)$: $U(\beta)=U(k)|_{e^{ik}\rightarrow \beta}$.
In the following, we first demonstrate the condition (in GBZ) for such a symmetry, and then extract the location of the non-Bloch exceptional points.

In our model, the Bloch Floquet operator in the BZ is
\begin{align}
U(k)=\exp\left(-i\frac{\theta_{1}}{2}\sigma_{y}\right)\exp\left(i\frac{k}{2}\sigma_{z}\right)
\exp\left(-i\frac{\theta_{2}}{2}\sigma_{y}\right)\exp\left(\sigma_{z}\gamma\right)\exp\left(-i\frac{\theta_{2}}{2}\sigma_{y}\right)
\exp\left(i\frac{k}{2}\sigma_{z}\right)\exp\left(-i\frac{\theta_{1}}{2}\sigma_{y}\right),
\end{align}
which is obtained from the identifications $|x+1\rangle\langle x|\rightarrow e^{-ik}$, $|x-1\rangle\langle x|\rightarrow e^{ik}$ that lead to $S_1\rightarrow e^{-ik\frac{1-\sigma_z}{2}}$ and $S_2\rightarrow e^{ik\frac{1+\sigma_z}{2}}$.

To define the Floquet operator in the GBZ, we perform the replacement $e^{ik}\rightarrow \beta=|\beta|e^{ip}$
\begin{align}\label{Ubeta}
U(\beta)=&\exp\left(-i\frac{\theta_{1}}{2}\sigma_{y}\right)\exp\left[\frac{i}{2}\left(p-i\ln|\beta|\right)\sigma_{z}\right]\exp\left(-i\frac{\theta_{2}}{2}\sigma_{y}\right)\exp\left(\sigma_{z}\gamma\right)\exp\left(-i\frac{\theta_{2}}{2}\sigma_{y}\right)\nonumber\\&\times\exp\left[\frac{i}{2}\left(p-i\ln|\beta|\right)\sigma_{z}\right]\exp\left(-i\frac{\theta_{1}}{2}\sigma_{y}\right).
\end{align}

While $U(\beta)$ is a nonunitary $2\times 2$ matrix, we can see that it still satisfies  $\det{U(\beta)}=\det{U^{-1}(\beta)}=\det{U^{\dagger}(\beta)}=1$ according to Eq.~(\ref{Ubeta}). Therefore, the additional condition needed for the pseudo-unitarity of $U(\beta)$ is $\rm{Tr}\left[{U^{-1}(\beta)}\right]=\rm{Tr}\left[{U^{\dagger}(\beta)}\right]$. These two conditions ensure that the product and sum of the two eigenvalues are the same for $U^{-1}(\beta)$ and $U^{\dagger}(\beta)$, therefore the eigenvalues should be the same, and there exists a certain $\eta$ such that Eq.~(\ref{pseudo-Supp}) holds. After a straightforward calculation, we obtain
\begin{align}\label{TrU}
\text{Tr}\left[U^{-1}(\beta)-U^{\dagger}(\beta)\right]=2 i \cos\theta_{1}\sin p\left[\frac{1}{|\beta|}(\sinh\gamma-\cosh\gamma\cos\theta_{2})+|\beta|(\sinh\gamma+\cosh\gamma\cos\theta_{2})\right].
\end{align}
Inserting Eq.~(\ref{radius}) into Eq.~(\ref{TrU}), we have
%\begin{align}\label{TrUbeta}
%&{\rm Tr}[U^{-1}(\beta)]-{\rm Tr}[U^{\dagger}(\beta)]\nonumber\\
%=&e^{ip}\cos\theta_{1}[\frac{1}{2}{\rm sgn}(v)\sqrt{|uv|}-\frac{1}{2}{\rm sgn}(u)\sqrt{|uv|}]+e^{-ip}\cos\theta_{1}[-\frac{1}{2}{\rm sgn}(v)\sqrt{|uv|}+\frac{1}{2}{\rm sgn}(u)\sqrt{|uv|}]
%\end{align}
\begin{align}\label{TrUbeta}
{\rm Tr}\left[U^{-1}(\beta)-U^{\dagger}(\beta)\right]=2i\cos\theta_{1}\sin p
\left[-\sqrt{|uv|}{\rm sgn}(u)+\sqrt{|uv|}{\rm sgn}(v)\right],
\end{align}
where we have used the shorthand notations
\begin{equation} u=\cosh\gamma\cos\theta_{2}-\sinh\gamma, \quad v=\cosh\gamma\cos\theta_{2}+\sinh\gamma. \end{equation}
%\begin{align}\label{TrUbeta}
%&u=\cosh\gamma\cos\theta_{2}-\sinh\gamma\\
%&v=\cosh\gamma\cos\theta_{2}+\sinh\gamma
%\end{align}
It follows that
\begin{align}\label{TrUlast}
{\rm Tr}\left[U^{-1}(\beta)-U^{\dagger}(\beta)\right]=\begin{cases}
0 & |\cos\theta_{2}|>|\tanh\gamma|\\
\pm4i\cos\theta_{1}\sin p\sqrt{|\cosh^{2}\gamma\cos^{2}\theta_{2}-\sinh^{2}\gamma|} & |\cos\theta_{2}|<|\tanh\gamma|
\end{cases}.
\end{align}
Therefore, we have shown that a necessary condition for $U(\beta)$ being pseudo-unitary is
\begin{equation} |\cos\theta_{2}|>|\tanh\gamma|. \label{condition} \end{equation}

We have then numerically checked that the parameter region with $|\cos\theta_{2}|>|\tanh\gamma|$ coincides with the exact \emph{PT} phase in the phase diagram Fig.~2(d) of the main text. Furthermore, we have checked that in the region with $|\cos\theta_{2}|>|\tanh\gamma|$, $U$ actually satisfies the strong version of pseudo-unitary condition~\cite{sMostafazadeh2}, namely that $\eta=\sum_n|\chi_n\rangle\langle\chi_n|$, where $|\chi_n\rangle$ is the $n$th left eigenstate of $U$, with $\langle\chi_n|U=\langle\chi_n|\lambda_n$ ($\lambda_n$ is the $n$th eigenvalue of $U$)~\cite{sMostafazadeh2}. This condition is equivalent to that $\eta$ can be decomposed as $\eta=OO^\dag$ with $O$ linear and invertible~\cite{sMostafazadeh2}.  This strong pseudo-unitarity guarantees $|\lambda_n|=1$ for all $n$, which translates to the complete reality of the quasienergy spectrum~\cite{sMostafazadeh2}. We have numerically checked that $\eta=\sum_n|\chi_n\rangle\langle\chi_n|$ indeed satisfies Eq.~(\ref{pseudo-Supp}). Therefore, we are able to conclude that the non-Bloch exceptional points are given by the expression $|\cos\theta_{2}|=|\tanh\gamma|$.

\begin{figure*}
\centering
\includegraphics[width=0.65\textwidth]{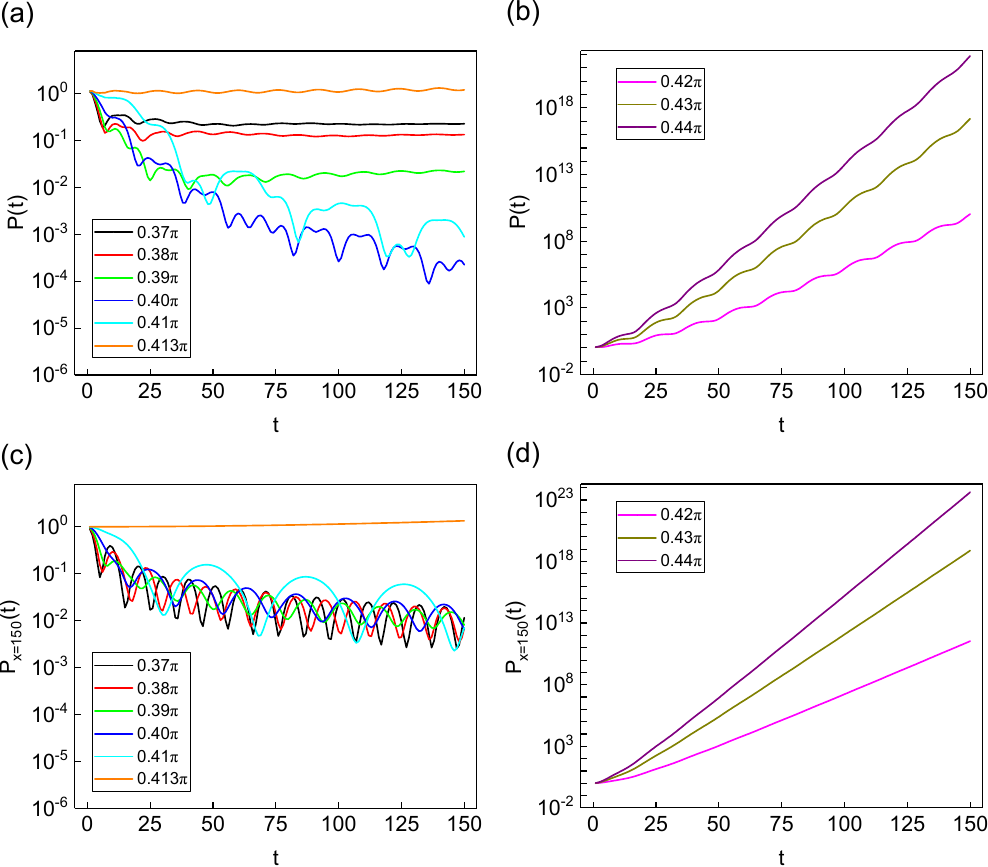}
\caption{ Numerical simulation of (a, b) the corrected probability $P(t)$ and (c, d) the corrected on-site probability $P_{x=150}(t)$ for an evolution of $150$ time steps. The coin parameters in (a, b) and (c, d) are the same as those in Fig. 2(e) and Fig. 3(e) of the main text, respectively. The initial states are $\ket{\psi(t=0)}=\ket{0}_x\otimes\ket{0}_\text{coin}$ in (a, b) and $\ket{\psi(t=0)}=\ket{150}_x\otimes\ket{0}_\text{coin}$ in (c, d), respectively. Here $\theta_2^R<0.413\pi$ indicates the exact \emph{PT} phase, while $\theta_2^R>0.413\pi$ indicates the \emph{PT} broken phase.
}
\label{fig:figS3}
\end{figure*}

\begin{figure*}
\centering
\includegraphics[width=0.65\textwidth]{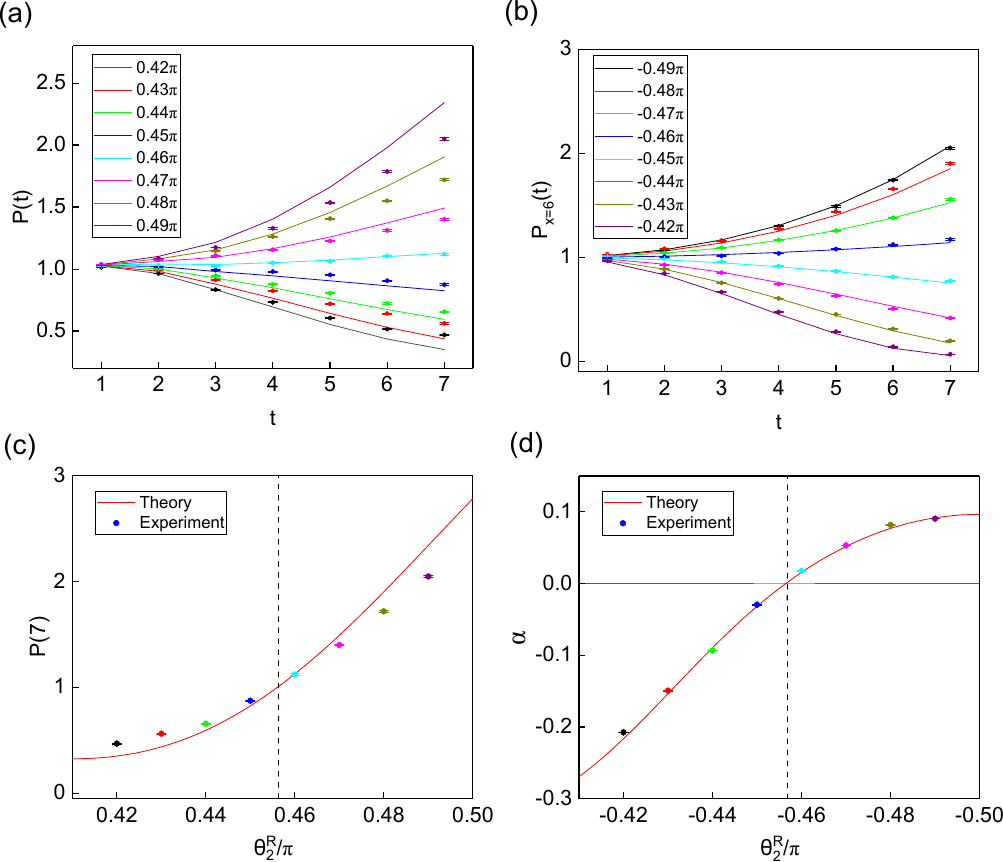}
\caption{Experimentally measured $P(t)$ and $P_{x=6}(t)$ (symbols) up to seven steps under the loss parameter $\gamma=0.1373$, together with theoretical predications (curves). (a) Coin parameters are $\theta_1^R=0.5625\pi$, $\theta_1^L=-0.0625\pi$, $\theta_2^L=0.75\pi$, and the initial state is $\ket{\psi(t=0)}=\ket{0}_x\otimes\ket{0}_\text{coin}$. (b) Coin parameters are $\theta_1^R=-0.4688\pi$, $\theta_1^L=-0.0625\pi$, $\theta_2^L=-0.9375\pi$ and the initial state is $\ket{\psi(t=0)}=\ket{6}_x\otimes\ket{0}_\text{coin}$. According to Eq.~(3) of the main text, the exceptional points are located at $\theta_2^R=0.4564\pi$ in (a), and $\theta_2^R=-0.4564\pi$ in (b), respectively. (c) $P(t=7)$ under the same parameter as those in (a). The red line is from numerical simulations of $7$-step evolutions, from which the exceptional point is predicted to be $\theta_2^R=0.4561\pi$. (d) Exponent $\alpha$ extracted from the data in (b). The red line is from numerical simulations, from which the exceptional point is predicted to be $\theta_2^R=-0.4564\pi$.
}
\label{fig:figS4}
\end{figure*}

\subsection{Numerical simulation for larger time steps}

While the experimental configuration limits the maximum number of achievable time steps, here we provide a numerical simulation for larger time steps under both schemes as discussed in the main text. As illustrated in Fig.~\ref{fig:figS3}, in the exact \emph{PT} phase [Figs.~\ref{fig:figS3}(a, c)], both the overall probability $P(t)$ in (a) and the on-site probability $P_{x=150}(t)$ in (b), exhibit fast oscillatory behavior enveloped by a slower decay in time, with the latter characterized by a negative fitting exponent $\alpha$. This is in sharp contrast to those in the \emph{PT}-broken phase [Figs.~\ref{fig:figS3}(b, d)], where both probabilities show exponential growth, with a positive fitting exponent $\alpha$. These features are consistent with our experimental observations under $7$-step time evolutions.

Furthermore, these results, the on-site probability in particular, are also consistent with calculations in~\cite{sLonghi2}, where the extracted exponent $\alpha$ is identified as the Lyapunov exponent in the long-time limit. It is shown that, in the exact \emph{PT} phase, $\alpha$ should approach zero from the negative side with increasing time steps; while in the \emph{PT}-broken phase, $\alpha$ should be positive. Such an understanding forms the theoretical basis of our second scheme.

\subsection{Experimental data with a different loss parameter}

To further demonstrate that the non-Bloch \emph{PT} transition predicted by Eq.~(3) of the main text should hold for other parameters, we perform additional experiments with an alternative loss parameter $\gamma=0.1373$. As shown in Fig.~\ref{fig:figS4}, all results support the relation in Eq.~(3), as are those in the main text.

\end{widetext}
\end{document}